\documentclass[aps,prb,twocolumn,floatfix,notitlepage]{revtex4-1}
\usepackage{amsmath,amsfonts,amssymb,amsthm,epsfig,array}
\usepackage{dsfont}
\usepackage{slashed}
\usepackage{graphics}
\usepackage{graphicx}
\usepackage{dcolumn}
\usepackage{float}
\usepackage{verbatim}
\usepackage{color}
\usepackage[dvipsnames]{xcolor}
\usepackage[hidelinks,colorlinks,linkcolor=blue,
citecolor=blue,urlcolor=blue]{hyperref}
\usepackage[titletoc,title]{appendix}
\usepackage{multirow}
\usepackage{bibentry}
\usepackage{tabularx}
\usepackage[mathscr]{euscript}
\usepackage{mathtools}
\allowdisplaybreaks
\usepackage{braket}

\DeclareMathOperator{\sgn}{sgn}
\DeclareMathOperator{\re}{Re}
\DeclareMathOperator{\im}{Im}
\DeclareMathOperator{\li}{Li}

\newcommand{\bsl}[1]{\boldsymbol{#1}}

\newcommand{\ii}{\mathrm{i}}

\renewcommand{\Re}{\mathop{\mathrm{Re}}}
\renewcommand{\Im}{\mathop{\mathrm{Im}}}

\newcommand{\eqnref}[1]{Eq.\,\eqref{#1}}
\newcommand{\figref}[1]{Fig.\,\ref{#1}}

\newcommand{\secref}[1]{Sec.\,\ref{#1}}
\newcommand{\appref}[1]{Appendix.\,\ref{#1}}
\newcommand{\refcite}[1]{Ref.\,[\onlinecite{#1}]}

\newcommand{\eq}[1]{\begin{equation} #1 \end{equation}}

\newcommand{\eqa}[1]{\begin{align}\begin{split} #1 \end{split}\end{align}}

\let\oldAA\AA
\renewcommand{\AA}{\text{\normalfont\oldAA}}

\newcommand{\eg}{{\emph{e.g.}}}

\newcommand{\bq}{\bar q}

\begin{document}
	\title{Presence versus absence of Two-Dimensional Fermi Surface Anomalies}
	\author{Donovan Buterakos}
	\thanks{These authors contributed equally to this work.}
	\affiliation{Condensed Matter Theory Center and Joint Quantum Institute, Department of Physics, University of Maryland, College Park, Maryland 20742-4111, USA}
	\author{DinhDuy Vu}
	\thanks{These authors contributed equally to this work.}
	\affiliation{Condensed Matter Theory Center and Joint Quantum Institute, Department of Physics, University of Maryland, College Park, Maryland 20742-4111, USA}
	\author{Jiabin Yu}
	\affiliation{Condensed Matter Theory Center and Joint Quantum Institute, Department of Physics, University of Maryland, College Park, Maryland 20742-4111, USA}
	\author{Sankar Das Sarma}
	\affiliation{Condensed Matter Theory Center and Joint Quantum Institute, Department of Physics, University of Maryland, College Park, Maryland 20742-4111, USA}
	\begin{abstract}
		We theoretically consider Fermi surface anomalies manifesting in the temperature dependent quasiparticle properties of two-dimensional (2D) interacting electron systems, comparing and contrasting with the corresponding 3D Fermi liquid situation.  
		In particular, employing microscopic many body perturbative techniques, we obtain analytically the leading-order and the next-to-leading-order interaction corrections to the renormalized effective mass for three distinct physical interaction models:  electron-phonon, electron-paramagnon, and electron-electron Coulomb coupling.  
		We find that the 2D renormalized effective mass does not develop any Fermi surface anomaly due to electron-phonon interaction, manifesting $\mathcal{O}(T^2)$ temperature correction and thus remaining consistent with the Sommerfeld expansion of the non-interacting Fermi function, in contrast to the corresponding 3D situation where the temperature correction to the renormalized effective mass has the anomalous $T^2 \log T$ behavior.  
		By contrast, both electron-paramagnon and electron-electron interactions lead to the anomalous $\mathcal{O}(T)$ corrections to the 2D effective mass renormalization in contrast to $T^2 \log T$ behavior in the corresponding 3D interacting systems.  
		We provide detailed analytical results, and comment on the conditions under which a $T^2 \log T$ term could possibly arise in the 2D quasiparticle effective mass from electron-phonon interactions.  
		We also compare results for the temperature dependent specific heat in the interacting 2D and 3D Fermi systems, using the close connection between the effective mass and specific heat.
	\end{abstract}
	\maketitle
	
	\begin{figure}[t]
		\centering
		\includegraphics[width=\columnwidth]{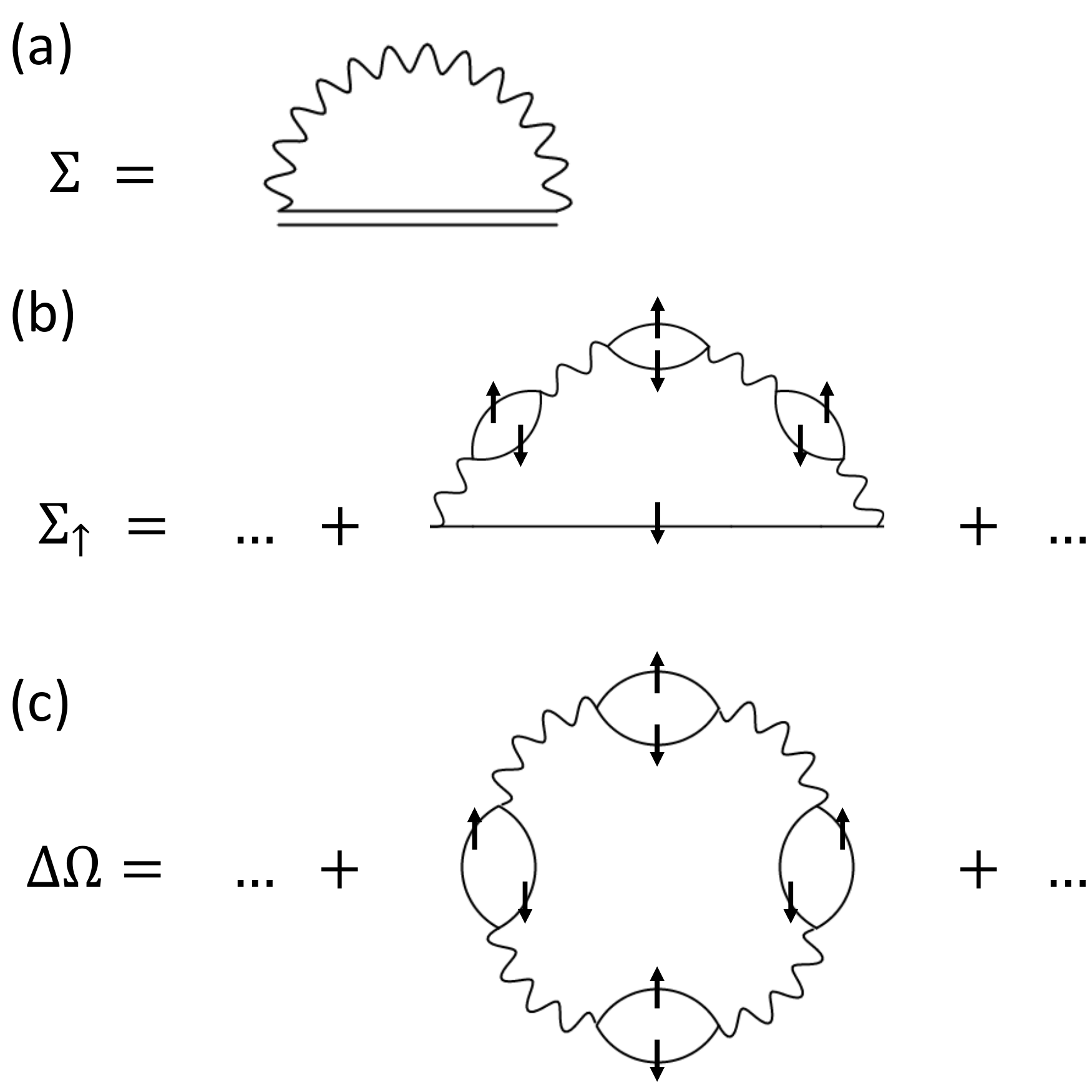}
		\caption{The diagrammatic contribution of interest to electron self-energy $\Sigma$ (a-b) and the total Free energy (c).
			In (a), the double straight lines stands for the exact propagator of electron, and the wavy line is the bare phonon propagator.
			In (b-c), the simple solid line is the bare electron propagator, and the wavy line corresponds to the paramagnon-mediated interaction.
			We only include the contribution from the spin flipping process, and the up and down arrows indicate the spin direction.
			Note that the anomalies show up already in the leading-order diagrams, and making the electron propagator self-consistent only adds higher-order terms which are parametrically smaller than the leading order anomalous results.
		}
		\label{fig:Feynman_Diagram}
	\end{figure}
	
	\section{Introduction}
	Fermi liquid theory is an astonishingly successful theory for interacting fermions asserting the perturbative preservation of the Fermi surface to all orders in the interaction, consequently leading to the existence of long-lived quasiparticles in the interacting system with one-to-one correspondence to the noninteracting particle-hole excitations~\cite{Abrikosov2012MethodsQFTStats}. 
	At low temperatures, therefore, interacting fermions are expected to manifest properties similar to a noninteracting Fermi gas except for a renormalization of the system parameters such as the effective mass, the Lande g-factor, the compressibility, etc.  
	In particular, one may expect that the Sommerfeld expansion remains applicable to the interacting system, implying that the leading-order thermal correction to various quasiparticle parameters should go as $\mathcal{O}(T^2)$ by virtue of the existence of the Fermi surface.  
	In particular, the interacting effective mass and specific heat should go as $\mathcal{O}(T^0) + \mathcal{O}(T^2)$ and $\mathcal{O}(T) + \mathcal{O}(T^3)$, respectively, following the Sommerfeld expansion results for the corresponding Fermi gas since thermal averaging over the Fermi surface should produce an $\mathcal{O}(T^2)$ next-to-the-leading-order thermal correction.  
	This, however, turns out not to be the case, and typically the next-to-the leading order temperature corrections to quasiparticle properties often manifest anomalous behavior not captured in the Sommerfeld expansion.  
	In the current work, we analytically study, using finite temperature many body perturbation theory, Fermi surface anomalies in 2D electron systems.
	
	The first theoretical report of an anomalous temperature dependence was by Eliashberg, who pointed out that the electron effective mass in an interacting 3D electron-phonon metallic system behaves as  $m^* [1 + \mathcal{O}(T^2 \log T)]$, where $m^*$ is the renormalized quasiparticle effective mass due to electron-phonon interaction~\cite{Eliashberg1963lowtempSpecificHeat}.
	The extra factor of $\log T$ is unanticipated in the Sommerfeld expansion, and is an interaction-induced Fermi surface anomaly.  
	Later on, it was realized that very similar Fermi surface anomalies involving a $T^2 \log T$ $(T^3 \log T)$ correction in the quasiparticle effective mass (specific heat) arise from 3D electron-paramagnon interactions as well~\cite{Berk1966FerroSpin,Doniach1966FerroLowTemp}.
	It therefore appears that such $\log T$ anomalous higher-order Fermi surface corrections arise in the quasiparticle effective mass renormalization due to fermion-boson interactions.  
	This is, however, not the case as was found out later when similar next-to-leading-order logarithmic temperature corrections were found to arise from pure electron-electron interactions also without involving any explicit bosonic terms in the Hamiltonian~\cite{Carneiro1975SpecificHeat}.
	These anomalous temperature corrections are thus intrinsic properties of the interacting Fermi surface which do not exist in the corresponding noninteracting Fermi gas and are outside the Sommerfeld expansion paradigm.
	
	In the current work we focus on 2D Fermi surface anomalies in interacting 2D electron system, comparing with the corresponding 3D situation.  
	There has been earlier work on 2D Fermi surface anomalies, focusing mostly on the effect of electron-electron interactions and often within the short-range interaction models~\cite{Chubukov2003NaanalyticCorrections,Chitov2001TempCorrections2D,Coffey1993NonanalyticBehavior2D}.
	There was an early finding that the Drude electrical resistivity of a 2D electron system interacting with random quenched charged impurities develops a linear-in-T correction instead of the naively expected $\mathcal{O}(T^2)$ thermal correction~\cite{Stern1980TemDepMobility,DasSarma1986FiniteTempScreening}.
	This served as a warning that 2D systems may be more singular than 3D systems manifesting nonanalytic $\mathcal{O}(T)$ corrections which are completely outside the Sommerfeld expansion paradigm which only obtains thermal corrections in powers of $T^2$ without allowing any appearance of a linear-in-$T$ term (or any odd powers of temperature).  
	This linear-in-$T$ interaction correction was later obtained for electron-electron interaction induced effective mass renormalization as well, establishing such nonanalytic linear in temperature corrections to be a generic feature of 2D Fermi surface anomalies~\cite{Chubukov2003NaanalyticCorrections,Galitski2004,Millis2006PRBSpecificHeat}.
	In the current work we further extend the story of 2D Fermi surface anomalies by considering the temperature corrections to the 2D electronic effective mass and specific heat renormalization arising from electron-phonon, electron-paramagnon, and electron-electron long-range Coulomb interactions all on the equal footing of leading-order many body theoretic calculations in the respective dynamical interactions (see \figref{fig:Feynman_Diagram}), comparing our 2D finding with the corresponding 3D logarithmic anomalies.  
	One interesting and unexpected finding is that the usual model of electron-phonon interaction does not lead to any Fermi surface anomaly in 2D in sharp contrast to the corresponding 3D situation with the $\log T$ correction originally discovered by Eliashberg \cite{Eliashberg1963lowtempSpecificHeat}.
	
	We provide here a brief summary of our results, with the theoretical and calculational details given in the subsequent sections.  
	Our work is completely analytical in nature.
	In this work, we discuss the Fermi surface anomalies for electron-phonon interaction, electron-paramagnon interaction, and electron-electron Coulomb interaction in two spatial dimensions.
	The theory is a leading-order many body perturbation theory (\figref{fig:Feynman_Diagram}), where the interaction (`wavy line') is the dynamical phonon or paramagnon interaction or the dynamically screened Coulomb interaction as the case may be.
	For the electron-phonon interaction, we find that although a nonanalytic $T^2\log T$ behavior naturally occurs in the 3D renormalized effective mass, the corresponding anomalous behavior in 2D for both linear and quadratic electron dispersion only arises from the higher-order momentum-dependence of the electron-phonon coupling and is therefore a rather small sub-leading effect at best.
	For the electron-paramagnon interaction, we find that an anomalous linear-in-$T$ term in the specitic heat $C_V/T$ appears in 2D for both linear and quadratic electron dispersion, in contrast to the $T^2\log T$ term in 3D.
	For the electron-electron Coulomb interaction, we verify the anomalous $T$ term in the 2D renormalized effective mass up to a similar $\mathcal{O}(T^2)$ order, in contrast to the $\mathcal{O}(T^2 \log T)$ renormalization of the 3D effective mass.
	
	The remaining of the paper is organized as the follows.
	We discuss the 2D/3D Fermi surface anomalies for the electron-phonon, electron-paramagnon, and electron-electron Coulomb interactions in \secref{sec:el-ph}, \secref{sec:el-mag}, and \secref{sec:el-el}, respectively.
	We conclude the paper in \secref{sec:sum}.
	An appendix provides details for some of the relevant integrals used in the main text.
	
	\section{Electron-phonon self energy}
	\label{sec:el-ph}
	In this section, we derive the finite temperature correction to effective mass of electrons arising from electron-phonon interactions.
	We consider the Debye model for the phonons, where a phonon's frequency is linearly proportional to its momentum and is limited by a cutoff known as the Debye frequency $\omega_D$.
	This cutoff can be represented in the dispersion relation with a Heaviside step function $\Theta$, so that the phonon dispersion becomes $E_{ph}(\bsl{q})=c|\bsl{q}|\,\Theta(q_D-|\bsl{q}|)$, where $q_D$ is the Debye momentum. 
	The second quantized Hamiltonian for this system is then:
	\eqa{
		H&=\sum_{\bsl{k},s} E_{\bsl{k}} c_{\bsl{k},s}^\dagger c_{\bsl{k},s}+\sum_{\bsl{q}}E_{ph}(\bsl{q})a_{\bsl{q}}^\dagger a_{\bsl{q}}\\
		&+\sum_{\bsl{k},\bsl{q}}g(\bsl{q})c_{\bsl{k},s}^\dagger c_{\bsl{k}+\bsl{q},s}\varphi_{\bsl{q}}\ ,
	}
	where $c_{\bsl{k},s}$ and $a^\dagger_{\bsl{q}}$ are creation operators for electrons and phonons, respectively, and $s=\uparrow\downarrow$ is for the spin index.
	In particular, $g(\bsl{q})$ is the electron-phonon coupling function, and $\varphi_{\bsl{q}}$ is related to the atom displacements as
	\eq{
		\varphi_{\bsl{q}}=\sqrt{E_{ph}(\bsl{q})}(a^\dagger_{\bsl{q}}+a_{-\bsl{q}})/2\ .
	}
	%
	%
	In this section, we use an arbitrary dispersion $E_{\bsl{k}}$ with the constraint that it must be isotropic, and thus $E_{\bsl{k}}$ does not depend on the direction of $\bsl{k}$. We define $k_F$ such that $E_{\bsl{k}}=\mu$ for $|\bsl{k}|=k_F$, where $\mu$ is the chemical potential, and we define $v_F$ as follows:
	\eq{
		v_F=\left.\frac{\partial E_{\bsl{k}}}{\partial |\bsl{k}|}\right|_{|\bsl{k}|=k_F}\ .
	}
	Note that $k_F$ and $v_F$ might be temperature-dependent due to their possible dependence on $\mu$. At zero temperature $k_F$ is the Fermi momentum and $v_F$ is the Fermi velocity. 
	Throughout the entire work, we adopt the unit systems in which $\hbar=k_B=1$.
	
	In \refcite{Buterakos2019ELPH}, the electron self-energy was calculated to one-loop order using a self-consistent approximation for $\bsl{q}$-independent $g(\bsl{q})$.
	In this section, we discuss a more general case where $g^2(\bsl{q})$ is a power function of $|\bsl{q}|$.
	\figref{fig:Feynman_Diagram}(a) shows the Feynman diagram for the self-energy within the self-consistent approximation, resulting in the following expression of the self-energy:
    \eqa{
		\label{eqn:sigma_wi}
		\Sigma(\omega_i)&=\sum_{\omega_j}\int\!\frac{d^d\bsl{k}}{(2\pi)^d}\;\frac{-g^2(\bsl{q})T }{\ii\omega_j-E_{\bsl{k}}+\mu-\Sigma(\omega_j)}\\
		&\times\frac{-c^2 |\bsl{q}|^2}{-(\omega_i-\omega_j)^2+c^2|\bsl{q}|^2}\ ,
	}
    where $-\ii\omega_i = (2i+1)\pi T$ is the fermionic Matsubara frequency. As was done in \refcite{Buterakos2019ELPH}, we change the variables of integration to $|\bsl{k}|$ and $|\bsl{q}|$, and perform the integration over $|\bsl{k}|$. To leading order in $q_D/k_F$, the only contribution to the $|\bsl{k}|$ integral is from the pole. The resulting expression is independent of the exact form of the dispersion $E_{\bsl{k}}$ aside from the dependence on $v_F$. Thus our calculation is applicable to systems with a linear or quadratic dispersion $E_{\bsl{k}}$. Then the self-energy expression simplifies to the following form:
	\eqa{
		\label{eqn:sigmanwi}
		\Sigma^{(n)}(\omega_i)&=\sum_{\omega_j}\int_0^{q_D}\!d|\bsl{q}|\;|\bsl{q}|^n\frac{-4g^2 T}{(2\pi)^2v_F}(-\ii \pi\sgn\im\omega_j)\\
		&\times\frac{-c^2 |\bsl{q}|^2}{-(\omega_i-\omega_j)^2+c^2|\bsl{q}|^2}\ .
	}
	In \eqnref{eqn:sigmanwi}, we have chosen $g^2(\bsl{q})=g^2 |\bsl{q}|^n$ for 2D and $g^2(\bsl{q})=2 g^2 |\bsl{q}|^{n-1}$ for 3D, where $g$ is the electron-phonon coupling constant, defining the interaction strength. We make this choice because the 3D self-energy expression is equal to the 2D expression with an additional factor of $|\bsl{q}|/2$. With this notation, the constant-$g(\bsl{q})$ self-energy expressions in \refcite{Buterakos2019ELPH} is given by $\Sigma^{(0)}$ in 2D and $\Sigma^{(1)}$ in 3D. Each value of $n$ defines a slightly different electron-phonon interaction model, and $n=0$ for 2D and $n=1$ for 3D are the usual physical electron-acoustic phonon interaction models in metals and semiconductors \cite{Grimvall}.
	
	By evaluating the sum and performing an analytical continuation on $\omega$, \eqnref{eqn:sigmanwi} is simplified and the retarded self-energy $\Sigma_R^{(n)}(\omega)$ is obtained as follows:
	\eqa{
		&\Sigma_R^{(n)}(\omega)=\frac{g^2}{2\pi^2v_F}\int_0^{q_D}d|\bsl{q}|\;|\bsl{q}|^n(c|\bsl{q}|)\bigg[-\pi \ii\coth\frac{c|\bsl{q}|}{2T}\\
		&\quad +\psi^{(0)}\Big(\frac{1}{2}+\ii\frac{c|\bsl{q}|-\omega}{2\pi T}\Big)-\psi^{(0)}\Big(\frac{1}{2}+\ii\frac{-c|\bsl{q}|-\omega}{2\pi T}\Big)\bigg]
	}
	where $\psi^{(n)}(z)$ is the polygamma function.
	In most cases the relevant energy scales for $\omega$ will be much smaller than the temperature of the system. Thus it is convenient to expand the result to first order in $\omega$, yielding:
	\eqa{
		\Sigma_R^{(n)}&=\frac{\omega g^2}{2\pi^2v_F}\int_0^{q_D}d|\bsl{q}|\;|\bsl{q}|^n\frac{\ii c|\bsl{q}|}{2\pi T}\bigg[\psi^{(1)}\Big(\frac{1}{2}-\ii\frac{c|\bsl{q}|}{2\pi T}\Big)\\
		&-\psi^{(1)}\Big(\frac{1}{2}+\ii\frac{c|\bsl{q}|}{2\pi T}\Big)\bigg]+\mathcal{O}\Big(\frac{\omega^2}{T^2}\Big).
	}
	Evaluating the integral gives the following:
	\begin{align}
		&\Sigma_R^{(n)}=\omega\frac{g^2}{2\pi^2v_F}\Big(\frac{2\pi T}{c}\Big)^{n+1}\nonumber\\&\times\Bigg((n+1)!\big(\ii^{n+1}+(-\ii)^{n+1}\big)\psi^{(-n-1)}\Big(\frac{1}{2}\Big)\nonumber\\&-\sum_{j=0}^{n+1}\frac{(n+1)!}{(n+1-j)!}\Big(\frac{\omega_D}{2\pi T}\Big)^{n+1-j}\nonumber\\&\times\bigg[\ii^j\psi^{(-j)}\Big(\frac{1}{2}+\frac{\ii\omega_D}{2\pi T}\Big)+(-\ii)^j\psi^{(-j)}\Big(\frac{1}{2}-\frac{\ii\omega_D}{2\pi T}\Big)\bigg]\Bigg).
		\label{eqn:srn}
	\end{align}
	Note that the system dimensionality (2D or 3D) enters only through the value of $n$ defining the momentum dependence of the electron-phonon interaction. This result can then be expanded for $T\ll\omega_D$ for any value of $n$ providing the low-temperature regime of interest. We give expansions for the first few values of $n$ to order $(T/\omega_D)^4$ below:
	\eqa{
		\Sigma_R^{(0)}&=\omega\frac{g^2}{2\pi^2v_F}\Big(\frac{\omega_D}{c}\Big)\bigg[-\!2+\frac{2\pi^2T^2}{3\omega_D^2}+\frac{14\pi^4T^4}{45\omega_D^4}\bigg]\\
		\Sigma_R^{(1)}&=\omega\frac{g^2}{2\pi^2v_F}\Big(\frac{\omega_D^2}{c^2}\Big)\bigg[-\!1+\frac{2\pi^2T^2}{3\omega_D^2}\log\frac{T}{\omega_D}\\
		&+\frac{\pi^2T^2}{\omega_D^2}\Big(1+\frac{2}{3}\log 4\pi-8\log A\Big)+\frac{7\pi^4T^4}{15\omega_D^4}\bigg]\\
		\Sigma_R^{(2)}&=\omega\frac{g^2}{2\pi^2v_F}\Big(\frac{\omega_D^3}{c^3}\Big)\bigg[-\frac{2}{3}-\frac{2\pi^2T^2}{3\omega_D^2}+\frac{14\pi^4T^4}{15\omega_D^4}\bigg]\\
		\Sigma_R^{(3)}&=\omega\frac{g^2}{2\pi^2v_F}\Big(\frac{\omega_D^4}{c^4}\Big)\bigg[-\frac{1}{2}-\frac{\pi^2T^2}{3\omega_D^2}+\frac{14\pi^4T^4}{15\omega_D^4}\log\frac{T}{\omega_D}\\
		&+\frac{\pi^4T^4}{\omega_D^4}\Big(\frac{7}{30}+\frac{16}{15}\log 2+\frac{14}{15}\log \pi-112\zeta'(-3)\Big)\bigg],
		\label{eqn:sigmavals}
	}
	where $A=1.282$... is Glaisher's constant, and $\zeta(z)$ is the Riemann zeta function. 
	Of particular interest are the $\log T$ terms in \eqnref{eqn:sigmavals} which appear in the expressions for $\Sigma_R^{(1)}$ and $\Sigma_R^{(3)}$. We also emphasize that odd powers of $T$ do not enter the low-temperature expression for the self-energy for any $n$, and even powers do show up, consistent with the Sommerfeld expansion, along with the anomalous $\log T$ terms for odd $n$ only.
	Next we will show that a logarithmic term appears in the expansion of $\Sigma_R^{(n)}$ if and only if $n$ is odd, and if so, it will have a prefactor of $T^{n+1}$. 
	
	To this end, we note that the only terms in the expansion of \eqnref{eqn:srn} that contain logarithms are the polygamma functions, in particular, their asymptotic expansions. 
	They can be expanded as series for large arguments such as:
	\begin{equation}
		\psi^{(0)}(z)=\log z-\sum_{k=1}^\infty\frac{B_k}{kz^k},
	\end{equation}
	where $B_k$ are the Bernoulli numbers with $B_1=+1/2$. 
	The asymptotic expansions of the other polygamma functions can be obtained up to a constant by integrating this series. 
	In particular, they produce the following terms with logarithms:
	\begin{equation}
		\psi^{(-j)}(z)=\sum_{k=0}^j\frac{(-1)^kB_k}{k!(j-k)!}z^{j-k}\log z+(\text{power series in $z$}).
		\label{eqn:polyexp}
	\end{equation}
	Let $a_m^n$ be the coefficient of the $T^m\log T$ term in the expansion of $\Sigma_R^{(n)}$. 
	Then from \eqnref{eqn:srn} and \eqnref{eqn:polyexp}, $a_m^n$ can be obtained, and after algebraic simplification it is found to be the following:
	\begin{align}
		a_m^n&=\omega\frac{g^2}{2\pi^2v_F}\Big(\frac{\pi \ii}{c}\Big)^{n+1}\big[1+(-1)^{n+1}\big]\delta_{m,n+1}\nonumber\\&\times\sum_{k=0}^{n+1}\binom{n+1}{k}B_k(-2)^k.
	\end{align}
	Thus we find that for $\Sigma_R^{(n)}$, there will be no $\log T$ terms if $n$ is even. If $n$ is odd, there will be exactly one logarithmic term at order $T^{n+1}\log T$.
	
	Using the expressions for the self-energy in \eqnref{eqn:sigmavals}, we calculate the temperature-dependent corrections to the effective mass $m^*$ according to the standard formula
	\eq{
		\frac{m^*}{m}=1-\left.\frac{\partial \Sigma^{(n)}_R(\omega)}{\partial \omega}\right|_{\omega\rightarrow0}\ .
		\label{eqn:mstarformula}
	}
	
	For the usual model of metallic electron-acoustic phonon interaction (with $n=0$ in 2D and $n=1$ in 3D) where $g^2(|\bsl{q}|)$ is independent of $|\bsl{q}|$ in \refcite{Buterakos2019ELPH}, we have
	\begin{equation}
		m^*_{\text{2D}}(T)=m\bigg[1+\frac{g^2 \omega_D}{c v_F}\bigg(\frac{1}{\pi^2}-\frac{T^2}{3\omega_D^2}\bigg)\bigg],
	\end{equation}
	\begin{align}
		m^*_{\text{3D}}(T)&=m\bigg[1+\frac{g^2 \omega_D^2}{c^2 v_F}\bigg(\frac{1}{4\pi^2}-\frac{1}{6}\frac{T^2}{\omega_D^2}\log\frac{T}{\omega_D}\nonumber\\&-\Big(\frac{1}{4}+\frac{\log 4\pi}{6}-2\log A\Big)\frac{T^2}{\omega_D^2}\bigg)\bigg].
		\label{eqn:mstar3delph}
	\end{align}
	We note that in 2D the correction to the effective mass due to the electron self-energy is of order $T^2$, whereas in 3D the correction is of order $T^2\log T$. 
	It means that the leading-order correction to the effective mass in 2D coincides with the Sommerfeld expansion, when the electron-phonon coupling is independent of the momentum.
	If $g^2(|\bsl{q}|)$ contains a linear-in-$|\bsl{q}|$ term  as $g^2(|\bsl{q}|)=g_0^2+g_1^2|\bsl{q}|$, then the 2D self-energy will include contributions from $\Sigma_R^{(1)}$ proportional to $g_1^2$.
	As a result, the total self energy, as well as the effective mass, will include a non-analytic $T^2\log T$ term, owing to the higher-order momentum dependence of $g^2(|\bsl{q}|)$. 
	On the other hand, if $g^2(|\bsl{q}|)$ only contains terms with even powers of $|\bsl{q}|$, then the 2D self energy will not possess $\log T$ terms, while the 3D self-energy does. We also mention an obvious feature of \eqnref{eqn:mstarformula}-\eqref{eqn:mstar3delph}, where the temperature-independent terms inside the square brackets provide the usual phonon-induced many-body electron effective mass renormalization at $T=0$~ \cite{Grimvall}.
	
	We note that Eliashberg explicitly considered $n=1$ for the 3D electron-phonon interaction, finding the failure of the Sommerfeld expansion in the next-to-the leading order which is $T^2 \log T$ for the effective mass and hence $T^3 \log T$ for the electronic specific heat in 3D metals\cite{Eliashberg1963lowtempSpecificHeat,Abrikosov2012MethodsQFTStats}. The same interaction model for 2D implies $n=0$, and hence no Fermi surface anomalies arising from electron-phonon coupling.  Thus, the existence or not of electron-phonon interaction induced Fermi surface anomaly is specifically model dependent and as such not a fundamental effect.  It is more of a coincidence in 3D, which does not appear in the standard model of electron-phonon interaction in 2D.  It is possible to generate higher order momentum dependence in $g(\bsl{q})$ even for $n=0$ in 2D by taking into account screening of the electron-phonon vertex by the electrons themselves, but such an effect can only introduce logarithmic anomalies at higher-order terms in the temperature expansion and are therefore irrelevant, being extremely small in magnitude as a high-order correction. Also, such a momentum expansion of any screened electron-phonon coupling explicitly necessitates the electronic Thomas-Fermi screening wavevector to be much larger than the phonon  Debye wavevector, a condition which most certainly does not apply to 2D systems.  As such, we can safely conclude that phonon-induced Fermi surface anomaly is absent (present) in 2D (3D). We comment that since the above calculations and arguments are independent of the form of the electron dispersion, the conclusions about the phonon-induced Fermi surface anomalies apply for both quadratic and linear (or any other) isotropic electron energy dispersion.
	
	We emphasize that the electron-phonon interaction we consider above is the electrons interacting  with bare phonons through a prescribed electron-phonon interaction, as was originally considered by Eliashberg in 3D finding that acoustic phonons by themselves, interacting with the electrons, lead to Fermi surface anomalies in 3D metals.  One may wonder what happens if the electron-phonon interaction is screened by the electrons themselves, which was not considered by Eliashberg since in 3D metals one thinks of the bare acoustic phonons as already arising from electronic screening of the ionic plasma modes within the standard jellium model (and thus, any additional screening would be over-counting).  If we use static screening~\cite{Grempel1982SpecificHeat2D}, \eg\  Thomas-Fermi screening, which is appropriate since the typical phonon frequency is much smaller than the typical electronic plasmon frequency, then the electron-phonon coupling, $g(\bsl{q})$, simply becomes a power series in $|\bsl{q}|$ as the screening function is expanded in a systematic momentum expansion.  This is equivalent to our power series form for $g(\bsl{q})$ considered above with $g^2(\bsl{q}) \sim |\bsl{q}|^ n$ for 2D and $g^2(\bsl{q}) \sim |\bsl{q}|^{n-1}$ for 3D.  On the other hand, dynamical screening of electron-phonon interaction is nothing other than the insertion of an infinite series of electron-hole bubbles dressing the phonon propagator (similar to what we show in \figref{fig:Feynman_Diagram}(b) for the paramagnon propagator), in which case, the phonon coupling problem becomes equivalent to the paramagnon problem (or the pure electron-electron interaction problem) considered below in \secref{sec:el-mag} (or \secref{sec:el-el}) of this article.  There would still be a difference between the 2D and the 3D cases in the sense that Fermi surface anomalies would rise at the zeroth (high) order in 3D (2D) arising from dynamical screening of phonons, and hence will be much weaker in 2D than in 3D as the effect will be suppressed by powers of $\omega_D/\omega_p$, where $\omega_p$ is the relevant electronic plasma frequency associated with the infinite series of electron-hole bubbles.  Since $\omega_p\gg\omega_D$ usually, the purely phonon-induced 2D Fermi surface anomalies would be weaker than the corresponding 3D anomalies.  The physics of electron-hole bubble induced Fermi surface anomalies is discussed below in \secref{sec:el-mag}-\ref{sec:el-el} in great details.
	
	\section{Electron-paramagnon Interaction}
	\label{sec:el-mag}
	
	In this section, we adopt a single-band model with Hubbard-type interaction $U$, whose Hamiltonian reads
	\begin{equation}
		\label{eq:H_paramagnon}
		H=\sum_{\bsl{k},s} E_{\bsl{k}}c_{s,\bsl{k}}^\dag c_{s,\bsl{k}} +  U \sum_{\bsl{q},\bsl{k},\bsl{k}'}c^\dagger_{\downarrow,\bsl{k}+\bsl{q}} c^\dagger_{\uparrow,\bsl{k}'} c_{\uparrow,\bsl{k}} c_{\downarrow,\bsl{k}'+\bsl{q}}\ .
	\end{equation}
	We consider a situation where electrons interact with paramagnons through a contact model interaction as in \eqnref{eq:H_paramagnon}. This interaction should be thought of as an effective short-range electron-paramagnon exchange interaction whose detailed form is not important for our consideration of Fermi surface anomalies.
	We set the electron dispersion $E_{\bsl{k}}$ to be given by \eqnref{eq:dispersion} for linear or quadratic dispersion as indicated.
    \eq{
		\label{eq:dispersion}
		E_{\bsl{k}}=\begin{cases}v_F|\bsl{k}|&\text{for linear dispersion}\\\frac{\bsl{k}^2}{2m}&\text{for quadratic dispersion}\end{cases}
	}
	%
    We consider both linear and quadratic dispersions in order to establish the generality of the results.
	
	To derive the self-energy and the specific heat arising from the paramagnon interaction, it is convenient to use the path integral formalism, in which the Hamiltonian \eqnref{eq:H_paramagnon} gives us the following action
	\eq{
		\label{eq:el-mag_action}
		S= \sum_{k,s=\uparrow\downarrow} G^{-1}(k) c^\dagger_{k,s} c_{k,s}+ \sum_{q} U \sigma^-_{q}\sigma^+_{q}\ .
	}
	Here $k=(\omega_n,\bsl{k})$ with $-\ii\omega_n=(2n+1)\pi/\beta$ the fermionic Matsubara frequency, and $q=(\nu_m,\bsl{q})$ with $-\ii\nu_m= m 2\pi/\beta$ the bosonic Matsubara fraquency, and $\beta=1/T$. 
	$G$ in \eqnref{eq:el-mag_action} is the bare electron propagator
	\begin{equation}
		\label{eq:G_el_bare}
		G(k)=\frac{1}{\omega_n -  E_{\bsl{k}} + \mu} 
	\end{equation}
	with the chemical potential $\mu$ , and 
	\begin{equation}
		\sigma^-_q = \sum_{k}c^\dag_{\downarrow,k+q}c_{\uparrow,k}
	\end{equation}
	stands for the paramagnon (spin-flipping) field.
	
	\subsection{Paramagnon contribution to effective mass}
	
	In this subsection, we consider the quadratic dispersion in~\eqnref{eq:dispersion} and review the paramagnon contribution to the effective mass in 2D and 3D. 
	To solve for the effective mass correction, we should first derive the self-energy of electrons.
	Although the Hubbard-type interaction in \eqnref{eq:el-mag_action} has various types of contributions to the electron self-energy, we focus on the spin-flipping contribution as shown in \figref{fig:Feynman_Diagram}(b), which is the transverse-spin-fluctuation contribution introduced in \refcite{Brinkman1968SFSH} for 3D calculations.
	The underlying reason for making such a choice is two-fold: (i) we believe that the main correction due to electron-paramagnon scattering is included in \figref{fig:Feynman_Diagram}(b), up to some $\mathcal{O}(1)$ factor; (ii) we are concerned only with the scaling behavior of the correction with respect to $T$ instead of making precise quantitative predictions which would require an accurate knowledge of $U$ anyway.
	
	Since we consider the paramagnetic phase where the bare electron Green function \eqnref{eq:G_el_bare} is spin-independent, the self-energy given by \figref{fig:Feynman_Diagram}(b) should also be spin-independent 
	\eq{
		\Sigma_{\uparrow}=\Sigma_{\downarrow}\equiv\Sigma\ .
	}
	Summing up the diagrams in \figref{fig:Feynman_Diagram}(b) and picking out the infinite-order collective effect~\cite{Brinkman1968SFSH}, we arrive at the following expression of the self-energy 
	\begin{equation}\label{eq:magonselfenergy}
		\Sigma(k) = -T\sum_{q}G(k-q)D(q)\ ,
	\end{equation}
	where $D(q)$ is the renormalized paramagnon propagator which will be addressed below (We note as an aside that \eqnref{eq:magonselfenergy} for the electron-paramagnon self-energy is formally the same as the corresponding expression given in \eqnref{eqn:sigmanwi} for the electron-phonon self-energy with $D$ in \eqnref{eq:magonselfenergy} being replaced by the corresponding phonon propagator in \eqnref{eqn:sigmanwi}).
	One can analytically continue the self-energy to real frequency and obtain the effective mass correction
	\begin{equation}
		\label{eq:effective_mass_correction}
		\begin{split}
			\frac{\Delta m}{m} &= \left.-\partial_\xi \text{Re}\Sigma_R(k_{\xi})\right|_{\xi=\mu}\\ &=-\int_{-\infty}^\infty \frac{d\nu}{2T\cosh^2(\nu/2T)} \\
			&\times   \int \frac{d^d \bsl{q}}{(2\pi)^d} \delta\left(\nu-v_F|\bsl{q}|\cos\theta-E_{\bsl{q}}\right)\text{Re}D_R(q)\ ,
		\end{split}
	\end{equation}
	where $k_{\xi}=(\xi,\bsl{k}_\xi)$, $|\bsl{k}_{\xi}|=\sqrt{2 m \xi}$, $d$ is the spatial dimension, and $\theta$ is the azimuthal angle of $\bsl{q}$.
	
	To use \eqnref{eq:effective_mass_correction}, the key step is to derive the retarded $D_R(q)$, which is the corresponding reducible interaction function and is given by
	\begin{equation}
		D_R(q)=\frac{U}{1-U\Pi^0_R(q)}\ ,
	\end{equation}
	where $\Pi_R^0$ is the non-interacting retarded electron spin-flip propagator.~\footnote{
		We note that $\Pi^0_R$ is different by nature from the conventional spin-preserving polarization operator; but as the electron propagator is spin-independent in the paramagnetic phase, the two diagrams have the same mathematical expressions. This allows us to express $\Pi^0_R$ by the Lindhard function.
	}
	Since the temperature-correction to $\Pi_R^0$ is of the order $\mathcal{O}(T^2)$ or higher for low $q$, we can neglect it and use the zero-temperature Lindhard expression~\cite{Lindhard1954} for $\Pi^0(q)$ in the limit $\bar{\nu} = \nu/(v_F k_F) \ll 1$ and $\bar{q} = |\bsl{q}|/k_F\ll 1$. 
	For the 3D case~\cite{Lindhard1954}, we have
	\begin{equation}
		\label{eq:Pi0_3D_mag}
		\begin{split}
			&\text{Re}\Pi_R^0(q)^{(3D)}=\rho_0\left(1-\frac{\bar q^2}{12}-\frac{\bar \nu^2}{\bar q ^2}\right)\\
			&\text{Im}\Pi_R^0(q)^{(3D)}= \frac{\pi \bar\nu\rho_0}{2\bar q}\Theta(\bar q -|\bar \nu|) 
		\end{split}
	\end{equation}
	with $\Theta(x)$ being the Heaviside function and $\rho_0$ being the density of states at the Fermi surface. 
	For the 2D case~\cite{Stern1967}, we have
	\begin{equation}
		\label{eq:Pi0_2D_mag}
		\begin{split}
			&\text{Re}\Pi_R^0(q)^{(2D)}=\rho_0\\
			&\text{Im}\Pi_R^0(q)^{(2D)}=\frac{\bar\nu \rho_0  \Theta(\bar q -|\bar \nu|) }{\sqrt{\bar q^2-\bar \nu ^2}}.
		\end{split}
	\end{equation}
	We set the cutoff of the momentum transfer to be $\Lambda k_F$ and consider the near-ferromagnetic situation (but still on the paramagnetic side) where $\alpha=1-\kappa=1-U\rho_0\ll 1$. 
	We can now calculate the effective mass correction in 2D and 3D.

	\subsubsection{3D Case}
	Let us first consider the $d=3$ case.
	After taking the integration over the solid angle in \eqnref{eq:effective_mass_correction}, the 3D effective mass correction becomes
	\begin{equation}
		\label{eq:3D_mass_Mag}
		\begin{split}
			\frac{\Delta m}{m}=\frac{-\rho_{F}}{4}\int \frac{d\nu}{2T\cosh^2(\frac{\nu}{2T})}\int_{|\bar\nu|}^{\Lambda} d\bar q \bar q \text{Re}D_R(q)
		\end{split}
	\end{equation}
	with
	\begin{equation}
		\text{Re}D_R(q)= \frac{U\left(\alpha+\frac{\bar q^2\kappa}{12} +\frac{\bar\nu^2\kappa}{\bq^2}\right)\bar q^2}{\left(\alpha+\frac{\bar q^2\kappa}{12} +\frac{\bar\nu^2\kappa}{\bq^2}\right)^2\bar q^2+\left(\frac{\pi\bar\nu \kappa }{2}\right)^2}.
	\end{equation}
    At low temperature $T \ll T_F =\mu$, expanding the result of the $\bar q$-integration in the limit $\nu\ll\alpha\ll 1$, we obtain up to $\mathcal{O}(\bar\nu^2)$
	\begin{equation}
		\begin{split}
			\int_{|\bar\nu|}^{\Lambda} d\bar q \bar q \text{Re}D_R(q)& =	\frac{6}{\rho_0}\log\left( \frac{\Lambda^2\kappa}{12\alpha} \right) \\
			& \qquad +\frac{\kappa^2(\pi^2\kappa+4\alpha)}{4\rho_0\alpha^3}\bar\nu^2\log|\bar\nu|.
		\end{split}
	\end{equation}
	This gives the effective mass correction as
	\begin{equation}
		\begin{split}
			\frac{\Delta m}{m} = &-3\log\left( \frac{\Lambda^2\kappa}{12\alpha} \right) 
			\\
			&-\frac{\pi^2\kappa^2(\pi^2\kappa+4\alpha)}{24\alpha^3}\left(\frac{T}{T_F}\right)^2\log\left(\frac{T}{T_F}\right)\ ,
		\end{split}
		\label{eq:3D_mass_Mag_final}
	\end{equation}
	coinciding with the $T^2\log T$ anomalous behavior presented in \refcite{Brinkman1968SFSH}.
	
	\subsubsection{2D Case}
	Now we turn to the $d=2$ case.
	The integration over solid angle in \eqnref{eq:effective_mass_correction} gives
	\begin{equation}
		\label{eq:2D_mass_Mag}
		\frac{\Delta m}{m} = \frac{-\rho_0}{\pi}\int \frac{d\nu}{2T\cosh^2\left(\frac{\nu}{2T}\right)}\int_{|\bar\nu|}^{\Lambda} \frac{d\bar q \bar q \text{Re}D_R(q)}{2\sqrt{\bar q^2-\bar\nu^2}}
	\end{equation}
	with
	\begin{equation}
		\text{Re}D_R(q) = \frac{U\alpha(\bar q^2-\bar \nu^2)}{\alpha^2(\bar q^2-\bar \nu^2)+(\kappa\bar\nu)^2}
	\end{equation}
	The integration over $\bar q$ up to $\mathcal{O}(\bar\nu^2)$ gives
	\begin{equation}
		\int_{|\bar\nu|}^{\Lambda} \frac{d\bar q \bar q \text{Re}D_R(q)}{2\sqrt{\bar q^2-\bar\nu^2}}=	\frac{\kappa\Lambda}{2\alpha\rho_0} - \frac{\pi\kappa^2}{4\rho_0\alpha^2}|\bar\nu|
	\end{equation}
	The corresponding effective mass correction is
	\begin{equation}
		\label{eq:2D_mass_Mag_final}
		\frac{\Delta m}{m} = -\frac{\kappa\Lambda}{2\pi\alpha} + \frac{\kappa^2\log 2}{2\alpha^2}\frac{T}{T_F}
	\end{equation}
	
	According to \eqnref{eq:3D_mass_Mag_final} and \eqnref{eq:2D_mass_Mag_final}, we have $T^2\log T$ term and $T$ term in the 3D and 2D effective mass as the next to the leading order temperature correction, respectively, both of which deviate from the Sommerfeld-like expansion (even powers of $T$) for the non-interacting Fermi gas.
	On first glance, it might be counter-intuitive that a linear-in-$T$ term arises in 2D because the integration with respect to $\bar q$ in \eqnref{eq:3D_mass_Mag} and \eqnref{eq:2D_mass_Mag} is even under $\nu\to -\nu$, which seemingly leads to even powers of $\nu$ after the integration and eventually to even powers of $T$ in the effective mass correction.
	However, the above logic only holds when the $\bar q$-integration gives an analytic function.
	In fact, the lower limit of the $\bar q$-integration is related to the singularity of the electron propagator, which coincides with the branch point of the paramagnon propagator, resulting in a non-analytic function that contain terms beyond the Sommerfeld-like expansion. This is the reason that such linear-in-$T$ 2D corrections are often called `nonanalytic' as they cannot arise from any thermal expansion of analytic functions which are guaranteed to produce even powers of $T$ in the expansion. Thus, the 2D Fermi surface anomaly arising from electron-paramagnon interaction is much stronger than that in 3D, being linear in $T$ compared with the $T^2 \log T$ anomaly in 3D. This is because the 2D Lindhard function is `more singular' than the 3D Lindhard function \cite{Stern1980TemDepMobility,DasSarma1986FiniteTempScreening}.
	
	\subsection{Paramagnon contribution to specific heat}
	
	The effective mass corrections in \eqnref{eq:3D_mass_Mag_final} and \eqnref{eq:2D_mass_Mag_final} imply that the specific heat $C_V/T$ should have $T^2\log T$ and $T$ corrections for 3D and 2D, respectively.
	In this subsection, we first verify this statement for quadratic dispersion and then show that the same Fermi surface anomalous corrections also occur for linear dispersion, defined by \eqnref{eq:dispersion}.
	
	Similar to above, we only include the spin-flipping contribution shown in \figref{fig:Feynman_Diagram}(b).
	Instead of the self-energy, we focus on the free energy, whose diagrams are given by connecting the two ends of \figref{fig:Feynman_Diagram}(b) by an extra fermion propagator [see \figref{fig:Feynman_Diagram}(c)].
	Such an operation would change the symmetry factor of each diagram in the summation, resulting in the following expression for the free energy
	\begin{equation}
		\label{eq:Delta_Free_Energy}
		\Delta \Omega = T\sum_{q} \log\left[1-U\Pi^0_R(q)\right]\ ,
	\end{equation}
	where $\Pi^0_R$ for quadratic dispersion has been shown in \eqnref{eq:Pi0_3D_mag} and \eqnref{eq:Pi0_2D_mag}.
	By analytic continuation, the free-energy shift can be evaluated by
	\begin{equation}
		\label{eq:DOmega_mag}
		\begin{split}
			\Delta \Omega &= \sum_{\bsl{q}}\frac{1}{\pi}\int_{-\infty}^{\infty}\frac{-d\nu}{e^{\beta\nu}-1}\text{Im}\left\{\log\left[1-U\Pi^0_R(q)\right] \right\} \\
		\end{split}\ ,
	\end{equation}
	and the correction to specific heat is just
	\eq{
		\frac{\Delta C_V}{T} = -\frac{\partial^2 \Delta\Omega}{\partial T^2}\ .
	}
	For compact representations of the results, we in the following will focus on $\Delta C_V/C_V^0$ instead of $\Delta C_V/T$, where $C_V^0=2\pi^2 T\rho_0/3$ is the specific heat of the non-interacting electron gas.
	
	\subsubsection{3D Case}
	In this part, we discuss the $d=3$ case, starting with the quadratic dispersion and then addressing the linear dispersion.
	
	For quadratic dispersion, we can substitute \eqnref{eq:Pi0_3D_mag} into \eqnref{eq:DOmega_mag} and obtain
	\begin{equation}
		\begin{split}
			\Delta \Omega = &\frac{k_F^3}{2\pi^3}\int\frac{d\nu}{e^{\beta\nu}-1} \int_{|\bar\nu|}^\Lambda d\bar q\bar q^2 I(\bar\nu,\bar q),
		\end{split}
	\end{equation}
	where the integration limit is obtained from the condition of non-zero $\Im\Pi^0_R$. We use $\tan^{-1}x\approx x - x^3/3$ to get
	\begin{equation}
		I(\bar\nu,\bar q) \approx \frac{\kappa\pi\bar\nu}{2\bar{q} \left(\alpha+ \bar q^2 \kappa/12\right)} - \frac{\kappa^2\pi(12\alpha+\pi^2\kappa)\bar\nu^3}{24\bar q^3 \alpha^3}\ .
	\end{equation}
	Straightforward integration gives $\Delta\Omega$ up to $\mathcal{O}(T^4)$ as
	\begin{equation}
		\begin{split}
			\Delta\Omega = &\frac{mk_F}{2}\ln\left(\frac{\Lambda^2\kappa}{12\alpha}\right)T^2\\
			&\quad+\frac{\kappa^2\pi^2(12\alpha+\pi^2\kappa)m^3}{360k_F^3}T^4\log\left( \frac{T}{T_F} \right),
		\end{split}
	\end{equation}
	resulting in the correction to the specific heat as
	\begin{equation}
		\begin{split}
			\frac{\Delta C_V}{C_V^0} = &-3\log\left(\frac{\Lambda^2\kappa}{12\alpha}\right) \\ 
			& \quad -\frac{\kappa^2\pi^2(12\alpha+\pi^2\kappa)}{40\alpha^3}\left(\frac{T}{T_F}\right)^2\log\left(\frac{T}{T_F}\right). 
		\end{split}
	\end{equation}
	Compared with \eqnref{eq:3D_mass_Mag_final}, the specific heat correction for quadratic dispersion also has $T^2\log T$ term up to a $\mathcal{O}(1)$ coefficient change, coinciding with \refcite{Brinkman1968SFSH}. Both effective mass and specific heat thus have the same 3D $T^2 \log T$ anomalous temperature dependence as expected since specific heat should be roughly proportional to the effective mass.
	
	For the linear dispersion, the real and imaginary parts of the polarization operator have similar forms to the quadratic dispersion case (see \appref{app:elmag_LD}).
	By performing exactly the same steps, we obtain the expression of the specific heat correction in the 3D linear dispersion case as
	\begin{equation}
		\begin{split}
			\frac{\Delta C_V}{C_V^0} = &-\frac{9}{2}\log\left(\frac{\Lambda^2\kappa}{12\alpha}\right) \\ 
			& \quad -\frac{\kappa^2\pi^2(12\alpha+\pi^2\kappa)}{10\alpha^3}\left(\frac{T}{T_F}\right)^2\log\left(\frac{T}{T_F}\right)\ , 
		\end{split}
	\end{equation}
	which also contains the nonanalytic $T^2\log T$ term similar to the 3D quadratic dispersion case.
	
	We mention that the Fermi surface for a 3D linear dispersion typically has a nonzero Chern number~\cite{Wan2011WSM}.
	As a result, the Fourier transformation of $c^\dagger_{\bsl{k}}$ in \eqnref{eq:H_paramagnon} is not localized in the real space, and neither is the Hubbard-type interaction in \eqnref{eq:H_paramagnon}.
	Nevertheless, we are still allowed to use \eqnref{eq:H_paramagnon} since the interaction can be viewed as a low-energy channel of a localized interaction that involves high-energy modes. Neglect of topology in considering low temperature anomalous renormalization correction is therefore justified.
	
	\subsubsection{2D Case}
	In this part, we discuss the renormalized specific heat for the $d=2$ case, again starting with the quadratic dispersion and then addressing the linear dispersion.
	
	For quadratic dispersion, we can combine \eqnref{eq:Pi0_2D_mag} with \eqnref{eq:DOmega_mag} to derive
	\begin{equation}
		\begin{split}
			\Delta\Omega &= \frac{k_F^2}{4\pi^2}\int \frac{d\nu \bar\nu^2}{e^{\beta\nu}-1}\int_{1}^{X} dx \tan^{-1}\left(\frac{\kappa}{\alpha\sqrt{x-1}}\right)\\
			& = \frac{ m \kappa\Lambda}{12 \alpha} T^2 - \frac{m^2\kappa^2\zeta(3)}{4\pi \alpha^2 k_F^2}T^3 + \mathcal{O}(T^4),
		\end{split}
	\end{equation}
	where $x=\bar q^2/\bar\nu^2$ and the integration limit $X=\Lambda^2/\bar\nu^2$. 
	Then, the corresponding specific heat correction is
	\begin{equation}
		\frac{\Delta C_V}{C_V^0} = -\frac{\kappa\Lambda}{2\pi\alpha} + \frac{9\zeta(3)\kappa^2}{2\pi^2\alpha^2}\frac{T}{T_F}
		\label{eq_2Dheat}
	\end{equation}
	Similar to \eqnref{eq:2D_mass_Mag_final}, the non-analytic liner-in-$T$ term also appears in the 2D specific heat correction for quadratic dispersion, as expected.
	
	In the 2D linear dispersion system, the real and imaginary parts of the polarization of the operator are exactly the same as in 2D quadratic dispersion case (see \appref{app:elmag_LD}). As a consequence, the specific heat correction in this case is simply \eqnref{eq_2Dheat}, which has a nonanalytic $T$ term. Thus, both effective mass and specific heat have nonanalytic $\mathcal{O}(T)$ corrections in 2D, independent of electron energy dispersion, arising from paramagnon renormalization in contrast to the 3D situation of $\mathcal{O}(T^2 \log T)$ anomalies.
	
	We note that we consider only bubble diagrams (see \figref{fig:Feynman_Diagram}) in our theory because we are taking into account scattering between opposite spin electrons as our system is paramagnetic.  For a spin-polarized (or spinless) system~\cite{How2018Nonanalytic}, one must consider the ladder diagrams, but all that does is to simply modify the nonuniversal numerical coefficient in front of the nonanalytic correction.  We also note that we use the usual Stoner-Hubbard-Mott model assuming the paramagnon coupling strength to be a constant on-site interaction, defined by $U$, but considering a more complicated interaction term does not change the leading-order Fermi surface anomalies obtained in the current work --- such a complicated momentum dependence in the paramagnon coupling only leads to higher-order corrections which are parametrically smaller.
	
	\section{Electron-Electron Coulomb Interactions}
	\label{sec:el-el}
	
	In this section, we briefly discuss the contribution to the effective mass from electron-electron Coulomb interactions. 	%
	In \refcite{Galitski2004}, the electron self-energy $\Sigma_R(\omega,T)$ was calculated at $\omega=0$ within the random phase approximation, which is exact in the high density and low temperature limits $T/T_F\ll r_s\ll1$, where $r_s$ is the dimensionless Coulomb coupling parameter. 
	This was used to show that in 2D, the temperature-dependent correction to effective mass is linear in $T$, and in 3D it is of order $T^2\log T$. 
	The 2D self-energy calculation was revisited in \refcite{LiaoPRB2020}, where $\Sigma_R(\omega,T)$ was calculated for arbitrary relative energy and temperature scales. 
	The corresponding Feynman diagrams for the self-energy are the same as \figref{fig:Feynman_Diagram}(a), except that here the double solid black line represents the bare electron propagator and the wavy line stands for the dynamically screened Coulomb interaction.
	Specifically, the following expression for the one-loop zero-temperature polarization bubble $\Pi_0$ is used \cite{Stern1967}:
	\begin{align}
	    \Pi_0(\bsl{q},\omega)&=-2\ii\int\frac{d^2\bsl{k}d\varepsilon}{(2\pi)^3}G_0(\bsl{k},\varepsilon)G_0(\bsl{k}+\bsl{q},\varepsilon+\omega)\nonumber\\
	    &=\frac{m\ii}{\pi |\bsl{q}|}\Bigg(\sqrt{2m\mu-\Big(\frac{m\omega}{|\bsl{q}|}+\frac{|\bsl{q}|}{2}+\ii0\Big)^2}\nonumber\\&\qquad-\sqrt{2m\mu-\Big(\frac{m\omega}{|\bsl{q}|}-\frac{|\bsl{q}|}{2}+\ii0\Big)^2}+\ii|\bsl{q}|\Bigg)
	\end{align}
	where $G_0$ is the electron Green's function, and the $\ii0$ term determines how to take the branch cut of the square root. This expression is for a quadratic electron dispersion, as defined in \eqnref{eq:dispersion}. By summing the Dyson series, the dynamically screened interaction potential $V_R(\bsl{q},\omega)$ is obtained:
	\begin{equation}
	    V_R(\bsl{q},\omega)=\Big[V_0(\bsl{q})^{-1}-\Pi_0(\bsl{q},\omega)\Big]^{-1}
	\end{equation}
	where $V_0(\bsl{q})=2\pi e^2/|\bsl{q}|$ is the bare Coulomb interaction potential. Then the self-energy is calculated from the Feynman diagram shown in \figref{fig:Feynman_Diagram}(a):
	\begin{equation}
        \Sigma(\bsl{k},\varepsilon)=-T\sum_\omega\int\!\frac{d^2\bsl{q}}{(2\pi)^2}\,\frac{V_R(\bsl{q},\omega)}{\varepsilon-\omega-E_{\bsl{k}-\bsl{q}}}
    \end{equation}
	In \refcite{LiaoPRB2020}, the sum over Matsubara frequencies is evaluated via contour integration, which requires calculating the real and imaginary parts separately, and introduces $\tanh\varepsilon/T$ terms. In order to evaluate the $\bsl{q}$ integral, the integrand is expanded to second order in $r_s$, and is also expanded in the quantity $\omega/(E_Fr_s)$. The resulting expression for the real part of the on-shell self-energy is:
	\begin{align}
	\label{eq:self_energy_LiaoPRB2020}
	    &\re\Sigma_R(\varepsilon)=\varepsilon\frac{r_s}{\sqrt{2}\pi}\log\frac{2\sqrt{2}}{r_s}\nonumber\\&-\frac{1}{8}\frac{T^2}{E_F}\big[\li_2(-e^{-\varepsilon/T})-\li_2(-e^{\varepsilon/T})\big]\nonumber\\&+\frac{5T^2\varepsilon}{48\sqrt{2}\pi r_s E_F^2}\Big(\pi^2+\frac{\varepsilon^2}{T^2}\Big)\log\frac{r_s E_F}{T}\nonumber\\&-\frac{32-10\gamma-25\log 2}{96\sqrt{2}\pi}\Big(\pi^2+\frac{\varepsilon^2}{T^2}\Big)\frac{T^2\varepsilon}{r_s E_F^2}\nonumber\\&-\frac{5T^3}{8\sqrt{2}\pi r_s E_F^2}\big[\partial_3\li_3(-e^{-\varepsilon/T})-\partial_3\li_3(-e^{\varepsilon/T})\big]
	\end{align}
	where $\gamma=0.577$... is Euler's constant, $\li_s(z)$ is the polylogarithm function, and $E_F$ is the Fermi energy, equal to $\mu$ at $T=0$.
	Using \eqnref{eqn:mstarformula} and the self-energy expression from \refcite{LiaoPRB2020} (\eqnref{eq:self_energy_LiaoPRB2020}), we calculate the 2D effective mass to be:
	\begin{align}
		&m^*_{\text{2D}}(T)=m\bigg[1-\frac{r_s}{\sqrt{2}\pi}\log\frac{2\sqrt{2}}{r_s}+\frac{\log 2}{4}\frac{T}{E_F}\nonumber\\&+\frac{5\pi}{48\sqrt{2}r_s}\frac{T^2}{E_F^2}\log\frac{T}{E_Fr_s}\nonumber\\&+\frac{T^2}{E_F^2r_s}\bigg(\frac{\pi(32-10\gamma-25\log 2)}{96\sqrt{2}}+\frac{5(6\zeta'(2)+\pi^2\log 2)}{48\sqrt{2}\pi}\bigg)\bigg].
	\end{align}
	Thus, there is a linear-in-$T$ correction to the effective mass in two dimensions which is independent of $r_s$, verifying the results first obtained in \refcite{Galitski2004}. This is in contrast to the corresponding 3D case where electron-electron interactions lead to an $\mathcal{O}(T^2 \log T)$ correction to the interacting effective mass and specific heat \cite{Galitski2004,Carneiro1975SpecificHeat}. It may be useful to point out that the Hartree-Fock approximation for the electron self-energy, where the electron-electron interaction in Fig.~\ref{fig:Feynman_Diagram} is taken as just the bare Coulomb interaction without any dynamical screening, leads to a singular result for the effective mass renormalization and the specific heat.  In particular, the Hartree-Fock renormalized 2D and 3D effective mass vanishes logarithmically in the leading order as $-1/(\log T)$ as $T$ goes to zero whereas the corresponding renormalized specific heat manifests an anomalous $–T/\log T$ behavior already in the leading order.  Such a leading-order singular behavior of effective mass and specific heat is inconsistent with experimental observations, and it is well-established that the Hartree-Fock approximation fails for Coulomb interaction in the calculation of the renormalized effective mass and specific heat.
	
	We note that one can ask whether electron-plasmon interactions would lead to Fermi surface anomalies similar to what we find for electron-paramagnon or dynamically screened electron-phonon interactions since plasmons are also bosons. The answer is affirmative since the plasmons are the direct consequence of the dynamical screening of the long range electron-electron interactions, and the bubble diagrams considered in \figref{fig:Feynman_Diagram} directly produce the plasmon mode for Coulomb coupling. Thus our results of \secref{sec:el-el} may be considered to be the theory for Fermi surface anomalies arising from electron-plasmon interactions.
	
	\section{Conclusion}
	\label{sec:sum}
	
	We have theoretically considered renormalization corrections to the finite temperature effective mass (and specific heat) arising from electron-phonon, electron-paramagnon, and electron-electron interactions in 2D, comparing the results to the corresponding 3D situation.  
	The emphasis is on finding and understanding anomalous terms in the temperature expansion beyond the Sommerfeld expression of the free Fermi gas. We find that the standard metallic electron-phonon interaction does not produce any 2D Fermi surface anomalies, leading only to the usual $\mathcal{O}(T^2)$ and higher even powers of temperature corrections to the renormalized 2D effective mass in contrast to the 3D case where the next-to-the-leading-order temperature correction goes as $\mathcal{O}(T^2 \log T)$.  
	We show how 2D $\mathcal{O}(T^2 \log T)$ anomalies may arise from electron-phonon coupling through suitable modifications of the momentum dependence of the interaction strength, concluding that such anomalous temperature dependence induced by electron-phonon interaction is not an intrinsic Fermi surface anomaly, but is an accidental behavior arising from the details of the functional form of the electron-phonon coupling.  
	By contrast, the Fermi surface anomalies induced by electron-paramagnon and electron-electron interactions are intrinsic as they arise from the intrinsic behavior of the electron propagators and do not depend on the electron energy dispersion or fine-tuned functional forms of the interaction itself.  
	In particular, this intrinsic anomaly in 2D introduces a nonanalytic $\mathcal{O}(T)$ next-to-the-leading order correction to the 2D effective mass and specific heat in sharp contrast to the 3D situation where the anomaly has a $\mathcal{O}(T^2 \log T)$ correction.  
	The linear-in-$T$ temperature correction is particularly anomalous since Sommerfeld expansion disallows any odd power of temperature appearing in the thermal averaging over the Fermi surface.
	
	We note that the absence of Fermi surface anomalies associated with 2D electron-phonon interactions (unless the interaction is fine-tuned unphysically) demonstrates that such anomalous temperature dependence cannot be construed to be a simple manifestation of the singular effect of any generic fermion-boson interactions, and really arises from the non-analytic structures in the electronic response functions.  A fundamental conceptual  difference between the electron-phonon and electron-paramagnon (or electron-electron) anomalies is that the expansion parameter for the electron-phonon interaction is $T/T_D$, with $T_D$ being the phonon Debye temperature, whereas it is $T/T_F$, with $T_F$ being the electron Fermi temperature, for the electron-paramagnon and electron-electron interactions.  This clearly shows that the phonon-induced $\log T$ term in 3D obtained in \refcite{Eliashberg1963lowtempSpecificHeat} is accidental, and is not really a Fermi surface anomaly.  Therefore, the absence of any phonon-induced 2D anomalous terms is reasonable and unsurprising. 
	
	An important lesson is that the Fermi liquid theory with its one to one correspondence between the noninteracting and the interacting system applies only to the leading order, and the next-to-the -leading-order corrections may manifest anomalies not allowed in a noninteracting system as a matter of principle, and such anomalous corrections may very well differ qualitatively and unexpectedly between 2D and 3D, as exemplified by the presence/absence of the $\mathcal{O}(T^2 \log T)$ anomaly in 3D/2D arising from electron-phonon interactions, and the presence of nonanalytic $\mathcal{O}(T)$ correction in 2D in contrast to the $\mathcal{O}(T^2 \log T)$ correction in 3D arising from electron-paramagnon and electron-electron interactions.  In some sense, the 2D Fermi liquid is more fragile than the 3D Fermi liquid as reflected in the $\mathcal{O}(T)$ Fermi surface anomalies induced by paramagnon-electron and electron-electron  interactions even if the corresponding electron-phonon interaction induced anomaly is absent in 2D. 

	\section{Acknowledgement}
	This work is supported by the Laboratory for Physical Sciences.
	
	\bibliographystyle{apsrev4-1}
	\bibliography{reference}
	
	\appendix
	\section{Details on $\Pi^0_R$ in \eqnref{eq:Delta_Free_Energy}}
	\label{app:elmag_LD}

	In this appendix, we present an approach to quickly derive the leading terms in $\Pi^0_R$ in \eqnref{eq:Delta_Free_Energy} in the limit $\bar q \ll 1$ and $\bar \nu \ll 1$. 
	In particular, we demonstrate the similarity between the expressions of $\Pi^0_R$ for linear and quadratic dispersion.
	
	We first start with the expression in complex frequency $z$
	\begin{equation}
		\begin{split}
			\Pi^0(\bsl{q},z) & = \int \frac{d^n k}{(2\pi)^n} \frac{\Theta(\mu-E_{\bsl{k}+\bsl{q}/2})-\Theta(\mu-E_{\bsl{k}-\bsl{q}/2})}{z-E_{\bsl{k}+\bsl{q}/2}+E_{\bsl{k}-\bsl{q}/2}}
		\end{split}
		\label{eq_polarization}
	\end{equation}
	We use the expansion
	\begin{equation}
		\begin{split}
			&\Theta(\mu-E-\Delta) - \Theta(\mu-E)\\
			&\approx -\Delta \delta(\mu-E) -\frac{\Delta^2}{2}\delta'(\mu-E)-\frac{\Delta^3}{6}\delta''(\mu-E)
		\end{split}
	\end{equation}
	We only collect terms up to $\bar q^2$ and $(\bar z/\bar q)^2$ where $\bar z = z/(v_F k_F)$, thus we can rewrite Eq.~\eqref{eq_polarization} as
	\begin{equation}
		\Pi^0(\bsl{q},z) =\left[ I_1(E) - \frac{\partial I_2(E)}{\partial E}+\frac{\partial^2I_3(E)}{\partial E^2}\right] \bigg|_{E=\mu},
		\label{eq_expand}
	\end{equation}
	where
	\begin{equation}
		\begin{split}
			I_1(E_{\bsl{k}}) & = \rho(E_{\bsl{k}})\left(1- \frac{z}{\Omega_n} \int\frac{d\Omega}{z-A_1(k_F)|\bsl{q}|\cos\theta} \right)\\
			I_2(E_{\bsl{k}}) & =  \frac{\rho(E_{\bsl{k}})|\bsl{q}|^2 }{4\Omega_n} \int d\Omega A_2(|\bsl{k}|,\theta)\\
			I_3(E_{\bsl{k}}) & =	\frac{\rho(E_{\bsl{k}})|\bsl{q}|^2 }{24\Omega_n} \int d\Omega A_1(|\bsl{k}|)^2\cos^2\theta.
		\end{split}
	\end{equation}
	Here, $d\Omega$ and $\Omega_n$ are the solid angle increment and full solid angle in the $n-$dimensional space; $A_1$ and $A_2$ are given by expanding $E_{\bsl{k}+\bsl{q}}-E_{\bsl{k}}$
	\begin{equation}
		E_{\bsl{k}+\bsl{q}}-E_{\bsl{k}}\approx A_1(|\bsl{k}|)\cos\theta|\bsl{q}| + A_2(|\bsl{k}|,\theta)|\bsl{q}|^2\ ,
	\end{equation}
	where $A_1=|\bsl{k}|/m$ and $A_2=1/(2m)$ for the quadratic dispersion case, and $A_1=v_F$ and $A_2=v_F\sin^2\theta/(2|\bsl{k}|)$ for the linear dispersion case.

	\subsection{Three-dimensional space}
	In the 3D space, $(1/\Omega_3)\int d\Omega = (1/2)\int d\cos\theta$.
	\subsubsection{Quadratic dispersion}
	We first perform integration with the $\theta$ and obtain
	\begin{equation}
		\begin{split}
			I_1(E_{\bsl{k}}) & = \rho(E_{\bsl{k}})\left[1- \frac{\bar z}{2\bar q} \ln\left( \frac{\bar z + \bar q}{\bar z - \bar q} \right)\right]\\
			I_2(E_{\bsl{k}}) & = \frac{\mu\rho(E_{\bsl{k}}) }{4}\bar{q}^2\\
			I_3(E_{\bsl{k}}) & = \frac{\mu E_{\bsl{k}} \rho(E_{\bsl{k}}) }{18}\bar{q}^2\\
		\end{split}
	\end{equation}
	As a result,
	\begin{equation}
		\Pi^0(\bsl{q},z)=\rho_0\left[ 1 - \frac{\bar z}{2\bar q} \ln\left( \frac{\bar z + \bar q}{\bar z - \bar q} \right) -\frac{\bar q^2}{12}  \right].
	\end{equation}
	By analytic continuation from imaginary to real frequency $z \to \nu+\ii\delta$, we obtain 
	\begin{equation}
		\begin{split}
			&\Re \Pi_R^0(\bsl{q}, \nu)=\rho_0\left[ 1 - \frac{\bar \nu^2}{\bar q^2} -\frac{\bar q^2}{12}  \right] \\
			&\Im \Pi_R^0(\bsl{q}, \nu)=\frac{\pi\rho_0\bar \nu}{2\bar q} \Theta(\bar q-|\bar \nu|)
		\end{split}
	\end{equation}
	These expressions are identical to the expansion of the exact Lindhard function obtained from quadratic energy dispersion.

	\subsubsection{Linear dispersion}
	For the linear dispersion $I_1$ is identical to the quadratic one, the other two terms are
	\begin{equation}
		\begin{split}
			I_2(E_{\bsl{k}}) & = \frac{\mu^2\rho(E_{\bsl{k}}) }{12E_{\bsl{k}}}\bar{q}^2\\
			I_3(E_{\bsl{k}}) & = \frac{\mu^2\rho(E_{\bsl{k}}) }{72}\bar{q}^2\\
		\end{split}
	\end{equation}
	The imaginary-frequency polarization operator is given by
	\begin{equation}
		\Pi^0(\bsl{q},z)=\rho_0\left[ 1 - \frac{\bar z}{2\bar q} \ln\left( \frac{\bar z + \bar q}{\bar z - \bar q} \right) -\frac{\bar q^2}{18}  \right].
	\end{equation}
	The real and imaginary parts for when $z\to \nu+\ii\delta$ are
	\begin{equation}
		\begin{split}
			&\Re \Pi_R^0(\bsl{q}, \nu)=\rho_0\left[ 1 - \frac{\bar \nu^2}{\bar q^2} -\frac{\bar q^2}{18}  \right] \\
			&\Im \Pi_R^0(\bsl{q}, \nu)=\frac{\pi\rho_0\bar \nu}{2\bar q} \Theta(\bar q-|\bar \nu|)
		\end{split}
	\end{equation}
	The expression for the linear dispersion case is similar to the quadratic one except for the numerical factor of $\bar q^2$ due to the difference in the dispersion curvature.
	
	\subsection{Two-dimensional space}
	
	In the 2D space, $(1/\Omega_2)\int d\Omega=(2\pi)^{-1}\int d\theta$. For both the quadratic and linear dispersion cases,
	\begin{equation}
		I_1(E_{\bsl{k}})=\rho(E_{\bsl{k}})\left[ 1- \frac{\bar z }{F\left(\sqrt{\bar z^2-\bar q^2}, \bar z\right)} \right],
	\end{equation}
	where $F\left(x, y\right)=|\Re x|\sgn(\Re y) + \ii|\Im x|\sgn(\Im y)$.
	For the linear dispersion, $I_2\propto \rho(E)$ and $I_3\propto E\rho(E)$ but $\rho(E)=\text{const}$, so the contributions of $I_2$ and $I_3$ are canceled by the first and second derivatives, respectively (see \eqnref{eq_expand}).  For the linear dispersion, $I_2\propto \rho(E)/E$, $I_3\propto \rho(E)$ and $\rho(E)\propto E$, so the terms $I_2$ and $I_3$ do not contribute either. As a result, for both cases of dispersion,
	\begin{equation}
		\Pi^0(\bsl{q},z)=\rho_0\left[ 1 -\frac{\bar z }{F\left(\sqrt{\bar z^2-\bar q^2}, \bar z\right)} +\mathcal{O}(\bar q^3) \right].
	\end{equation}
	By analytic continuation, we obtain the retarded functions:
	\begin{equation}
		\begin{split}
			&\Re \Pi_R^0(\bsl{q}, \nu)=\rho_0 \\
			&\Im \Pi_R^0(\bsl{q}, \nu)=\frac{\rho_0\bar \nu}{\sqrt{\bar q^2-\bar\nu^2}} \Theta(\bar q-|\bar \nu|).
		\end{split} 
	\end{equation}   
	
\end{document}